# Error analysis for satellite gravity field determination based on two-dimensional Fourier methods

Lin Cai[1, 3], Zebing Zhou[2, 3], Houtse Hsu[4, 3], Fang Gao[2, 3], Zhu Zhu[2, 3], Jun Luo[2]

[1] Department of Electronics and Information Engineering, Huazhong University of Science and Technology, Wuhan 430074, China

[2] School of Physics, Huazhong University of Science and Technology, Wuhan 430074, China

[3] Institute of Geophysics, Huazhong University of Science and Technology, Wuhan 430074, China

[4] Institute of Geodesy and Geophysics (IGG), Chinese Academy of Sciences, Wuhan 430077, China

## Abstract

The time-wise and space-wise approaches are generally applied to data processing and error analysis for satellite gravimetry missions. But both the approaches, which are based on least-squares collocation, address the whole effect of measurement errors and estimate the resolution of gravity field models mainly from a numerical point of indirect view. Moreover, requirement for higher accuracy and resolution gravity field models could make the computation more difficult, and serious numerical instabilities arise. In order to overcome the problems, this study focuses on constructing a direct relationship between power spectral density of the satellite gravimetry measurements and coefficients of the Earth's gravity potential. Based on two-dimensional Fourier transform, the relationship is analytically concluded. By taking advantage of the analytical expression, it is efficient and distinct for parameter estimation and error analysis of missions. From the relationship and the simulations, it is analytically confirmed that the low-frequency noise affects the gravity field recovery in all degrees for the instance of satellite gradiometer recovery mission. Furthermore, some other results and suggestions are also described.

*Correspondence to*: Zebing Zhou

# 1 Introduction

The Earth's gravity field with unprecedented high accuracy and resolution is mapped by new dedicated satellite gravimetry missions during the last decades. The corresponding methods of error analysis, which determine science requirements and mission parameters, need be improved. Currently, these methods for Satellite-to-Satellite Tracking (SST) and gradiometry are mainly based on least-squares (LS) collocation theory, and basically divided into two types: the time-wise approach and the space-wise approach (see e.g. Rummel et al. 1993; Reigber et al. 2005; Pail et al. 2011). The former treats the observations as a sequence of discrete time series along the orbit, which leads to a relation between the Fourier and the spherical harmonic coefficients by a huge linear system of equations with a large number of unknowns, and be suitable to analyze the orbit characteristics. The latter treats the measurements as a function of spatial coordinates, which analyzes the data on a specific grid on the mean orbital sphere with spherical harmonics. Compared to the time-wise approach, the space-wise approach splits the linear system of equations into smaller systems, which can be solved more easily (Sneeuw 2003). Further details of the approaches can be also found in e.g. Schuh (1996), Klees et al. (2000), Tscherning (2001) and Pail and Plank (2002).

The solution to the equations based on LS has the optimum statistical properties, but both the time-wise and space-wise approaches address the effect of measurement errors and estimate the resolution of gravity field models mainly from a numerical point of indirect view. Especially the latter method has treated measurement errors as white noise and ignored the actual noise increases with $1/f$ at lower frequencies for a long time (Schrama 1990; Sneeuw et al. 2005). The essential reason is that spherical harmonics are represented in spatial domain but noise is measured in time domain, so the procedure to estimating the effect of noise is too complicated to be analytically described (see Migliaccio et al. 2004). The latest and incoming gravitational models with increasing accuracy and resolution make the computation more difficult since the computation become huge and serious numerical instabilities arise when degree/order of models get higher (Gruber et al. 2011).

For the reasons mentioned above, it is important to develop a direct and efficient procedure of error analysis for satellite gravity field determination. By using a method based on two-dimensional (2-D) Fourier theory, the direct relationship between the power spectral density (PSD) of the satellite measurement errors and coefficients of the Earth gravity potential is analytically concluded in this study. Moreover, with some assumptions, the simulations and discussions are presented.

## 2 Spectral analysis of the Earth's gravity field

The Earth's gravity field satisfies the Laplace equation and can be expanded into a series of spherical harmonics (Heiskanen and Moritz 1967). In this study the following spherical harmonics representation for the Earth's disturbance potential $T$ is adopted:

$$T(r,\theta,\lambda) = \frac{GM}{R} \sum_{l=2}^{\infty} \left(\frac{R}{r}\right)^{l+1} \sum_{m=0}^{l} \left(\bar{C}_{lm} \cos m\lambda + \bar{S}_{lm} \sin m\lambda\right) \bar{P}_{lm}(\cos\theta) \qquad (1)$$

where

$r, \theta, \lambda$      geocentric spherical coordinates (radius, co-latitude, longitude)

$R$      reference length (mean semi-major axis of the Earth)

$GM$      gravitational constant times mass of the Earth

$l, m$      degree, order of spherical harmonic

$\bar{P}_{lm}$      fully normalized Legendre functions

$\bar{C}_{lm}, \bar{S}_{lm}$      fully normalized potential coefficients

The potential coefficients $\bar{C}_{lm}$ and $\bar{S}_{lm}$ are the unknown coefficients that should be estimated from observations as the solution of the gravity field recovery problem. Observations $n$ are generally assumed as a functional of the disturbance potential $T$

$$n = LT \qquad (2)$$

where $L$ stands for a functional, which could be scalar, vectorial or tensorial, depending on the observation object $n$. More details of the functional for different space sensors can be found in Scharifi (2006) and Sneeuw (2000). For the sake of lucidity, the $T_{zz}$ component in GOCE framework will be dealt with, although the method can be applied to other kinds of observables.

The second-order radial derivative with respect to local north-oriented frame, whose z-axis is pointing upwards in geocentric radial direction, x-axis towards the north and the frame is right handed, is given by (see Koop 1993):

$$T_{zz}(r,\theta,\lambda) = \frac{GM}{R^3} \sum_{l=2}^{\infty} (l+1)(l+2) \left(\frac{R}{r}\right)^{l+3} \sum_{m=0}^{l} \left(\bar{C}_{lm} \cos m\lambda + \bar{S}_{lm} \sin m\lambda\right) \bar{P}_{lm}(\cos\theta) \quad (22)$$

As mentioned above, the measurement errors of the observations, i.e. gravity gradient component $T_{zz}$, determines the uncertainty of spherical coefficients $\bar{C}_{lm}$ and $\bar{S}_{lm}$. By applying the orthogonality property of spherical harmonics, the total power of the measurement errors of $T_{zz}$ over all degrees, denoted as $\sigma_{T_{zz}}$, can be obtained in the form (Ilk et al. 2004)

$$\sigma_{T_{zz}} = \frac{GM}{R^3} \sum_{l=2}^{\infty} \left(\frac{R}{r}\right)^{l+3} (l+1)(l+2) \sqrt{\sum_{m=0}^{l} \left(\sigma_{\bar{C}_{lm}}^2 + \sigma_{\bar{S}_{lm}}^2\right)} \quad (23)$$

For a specific value of $l$, the summation over $l$ at the right-hand side of Eq. (23) should be removed, and the errors power $\sigma_{T_{zz}}$ at the left-hand side is replaced with error degree power $\sigma_{T_{zz},l}$, which represents the error power introduced in the $l$-th degree. Then, the relationship can be written as

$$\sigma_l = \frac{\sigma_{T_{zz},l}}{\frac{GM}{R^3} \left(\frac{R}{r}\right)^{l+3} (l+1)(l+2)} \quad (24)$$

As mentioned in Section 2, if $T_{zz}$ error PSD $S_{zz}(f)$ is given, the error degree power $\sigma_{T_{zz},l}$ can be presented by integrating the PSD over several tiny frequency bands:

$$\sigma_{T_{zz},l} = \sqrt{\sum_j \int_{f_j-\Delta f/2}^{f_j+\Delta f/2} \left[S_{zz}(f)\right]^2 df} \quad (25)$$

With substitution of Eq. (25) into Eq. (24), one obtains:

$$\sigma_l = \frac{\sqrt{\sum_j \int_{f_j-\Delta f/2}^{f_j+\Delta f/2} \left[S_{zz}(f)\right]^2 df}}{\dfrac{GM}{R^3}\left(\dfrac{R}{r}\right)^{l+3}(l+1)(l+2)} \tag{26}$$

Consequently, the transfer function between $T_{zz}$ error PSD and error degree variance is obtained (cf. Cai et al. 2011), except that temporal frequencies corresponding to spherical harmonic coefficients have not be determined yet, which will be discussed in the next section.

## 3 Temporal frequencies determined by 2-D Fourier methods

The noise performed by space sensor is colored, especially for the low-frequency range, in which there are usually 1/$f$ characteristic noise (see ESA 1999; Pail 2010). So the temporal frequencies corresponding to the $l$-th degree spherical harmonics need to be located in the frequency axis, since the contribution of noise is different between high frequencies and low frequencies. In this section, the relationship between the spatial frequencies and the temporal frequencies is concluded by using 2-D Fourier methods. Over the mission duration, the measurements are sampled by the satellite at a certain velocity along the orbit, and so it is natural for us to try establishing the relationship by the velocity. In order to facilitate a better understanding, a simple case, which works on the assumption that there is only a single-frequency signal in both the latitude and longitude directions, will be first discussed, and then the temporal frequencies corresponding to spherical harmonics will be analyzed.

### 3.1 Temporal frequencies determination for a single-frequency signal in both the latitude and longitude directions

To get a better understanding of the relationship between temporal frequencies and spherical harmonics, we assume first that the single-frequency spatial signals, cos$q\theta$ and cos$p\lambda$ ($q$, $p$ are constants, and represent the spatial frequencies), are along the latitude and longitude directions, respectively. Additionally, this study bases on the theoretical hypothesis that an instrument orbits on a polar circular orbit and observations are acquired when the tracks are along the lines of longitude at equi-angular intervals (Sneeuw 1994; Sneeuw and Bun 1996), as shown in Fig. 1. If the measurements, taking a line as unit, are not arranged in the order of longitude, they need to be rearranged. This procedure has no effect on error characteristics under the postulate of stationary stochastic noises.

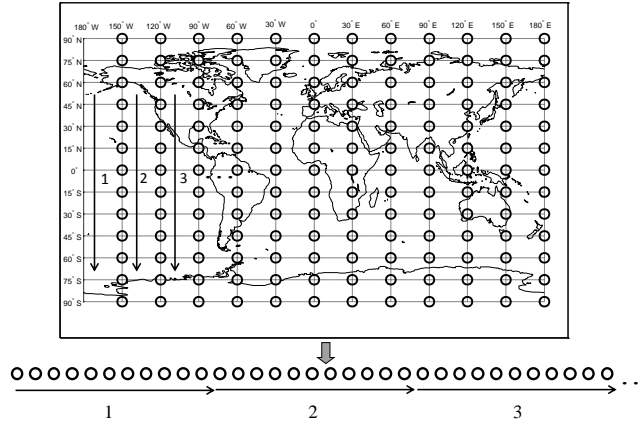

**Fig. 1.** The scheme for data regularly sampled on the Earth

It should be pointed out that all hypotheses are beneficial to the essence of the problem in spite of some difference in reality. To fulfill the 2-D sampling theorem, the minimum spatial sampling intervals in the latitude and longitude directions are $\pi/q$ and $\pi/p$, respectively. Accordingly, one obtains

$$\begin{cases} \cos q\theta, & \Delta\theta \leq \dfrac{\pi}{q} \\ \cos p\lambda, & \Delta\lambda \leq \dfrac{\pi}{p} \end{cases} \quad (11)$$

Due to Kepler's third law, the angular velocity $\omega$ (unit: rad/s) of the satellite is determined by $\omega = \sqrt{GM/r^3}$. The maximum degree of the gravity model, the number of the points in a longitude line and the number of the longitude lines are denoted as $L$, $K_\theta$ and $K_\lambda$, respectively. Then, the conditions $K_\theta \geq L$ and $K_\lambda \geq 2L$ are concluded by using 2-D sampling theorem, since the co-latitude $\theta$ goes from 0 to $\pi$ and the longitude $\lambda$ goes from 0 to $2\pi$. Based on the assumptions above mentioned, the angular velocity in the latitude direction $\omega_\theta$ equals $\omega$, and the velocity in the longitude direction $\omega_\lambda$ equals

$$\omega_\lambda = \frac{2\pi}{K_\lambda} \bigg/ \frac{\pi}{\omega} = \frac{2\omega}{K_\lambda} \quad (12)$$

and then the signals can be expressed in the time-dependent form:

$$\begin{cases} \cos q\theta = \cos(q\omega t) \\ \cos p\lambda = \cos\left(\dfrac{2\omega}{K_\lambda} pt\right) \end{cases} \quad (13)$$

The corresponding frequencies in the latitude and longitude directions are defined as $f_q$ and $f_p$, respectively, follows as

$$\begin{cases} f_q = q\dfrac{\omega}{2\pi} \\ f_p = \dfrac{2\omega}{K_\lambda}\cdot\dfrac{p}{2\pi} \end{cases} \quad (14)$$

Based on modulation theorem, the temporal frequency of the time series sampled by satellites $f_{q,p}$ is concluded by

$$f_{q,p} = \dfrac{\omega}{2\pi}\left(q + \dfrac{2p}{K_\lambda}\right) \quad (15)$$

This shows that temporal frequencies are determined by the constants $q$ and $p$. For this reason, the spatial frequencies in the latitude and longitude directions need to be computed. Additionally, the value of $f_{q,p}$ is mostly determined by $q$, since $K_\lambda$ is a large number in reality. In this subsection, the relationship between spatial domain and time domain is concluded, and lays a foundation for determining temporal frequencies corresponding to spherical harmonics.

### 3.2 Temporal frequencies determination for spherical harmonics

In this subsection, spherical harmonics are first expanded into sums of sine-cosine series, and then the corresponding temporal frequencies are determined base on the above mentioned result. The spherical harmonics $\bar{Y}_{lm}(\theta,\lambda)$ are defined in the following way:

$$Y_{lm}(\theta,\varphi) = \bar{P}_{lm}(\cos\theta)\begin{cases} \sin m\lambda \\ \cos m\lambda \end{cases} \quad \begin{pmatrix} m = 0,1,2,\cdots l \\ l = 0,1,2,3,\ldots \end{pmatrix} \quad (16)$$

where the normalized Legendre function $\bar{P}_{lm}(\cos\theta)$ is given by Colombo (1981):

$$\bar{P}_{lm}(\cos\theta) = \dfrac{(-1)(2l!)}{2^l l!(l-m)!}\sqrt{\dfrac{2(2l+1)(l-m)!}{(l+m)!}}\sin^m\theta \sum_{k=0}^{L(l,m)} a_k(l,m)\cos^{l-m-2k}\theta \quad (17)$$

with

$$L(l,m) = \begin{cases} (l-m)/2 & \text{if } (l-m) \text{ is even} \\ (l-m-1)/2 & \text{if } (l-m) \text{ is odd} \end{cases}$$

$$a_k(l,m) = \frac{(-1)^k (l-m)(l-m-1)\cdots(l-m-2k+1)}{2\cdot 4\cdots 2k(2l-1)\cdots(2l-2k+1)}$$

$\bar{P}_{lm}(\cos\theta)$ is expressed as a finite series in cosine or sine by Eq. (17). The elements related with longitude $\lambda$ of the spherical harmonics $\bar{Y}_{lm}(\theta,\lambda)$, while $\sin m\lambda$ and $\cos m\lambda$ are obviously Fourier series, so spherical harmonics can be expanded into sums of 2-D Fourier series in the domain $0 \leq \theta \leq \pi$, $0 \leq \lambda \leq 2\pi$.

From Eq. (17), it is also seen that the maximum degree of $\bar{P}_{lm}(\cos\theta)$ is $l$, since the highest frequency term corresponds to the highest frequency term $\sin^m\theta \cos^{l-m}\theta$. The degree of the terms decreases by a step size 2 when $k$ runs from 0 to $L(l, m)$, so $\bar{P}_{lm}(\cos\theta)$ has $(\frac{l}{2}+1)$ terms for $l$ even, and $(\frac{l-1}{2}+1)$ for $l$ odd. By using trigonometric product sum identities, the normalized Legendre function $\bar{P}_{lm}(\cos\theta)$ can be written as

$$\begin{cases} \bar{P}_{lm}(\cos\theta) = \sum_{i=0}^{N(l)} C_i \cos(l-2i)\theta & \text{if } m \text{ is even} \\ \bar{P}_{lm}(\cos\theta) = \sum_{i=0}^{N(l)} S_i \sin(l-2i)\theta & \text{if } m \text{ is odd} \end{cases} \quad (18)$$

with

$$N(l) = \begin{cases} \dfrac{l}{2} & \text{if } l \text{ is even} \\ \dfrac{l-1}{2} & \text{if } l \text{ is odd} \end{cases}$$

where $C_i$ and $S_i$ are the coefficients of Fourier series. Substitution of Eq. (18) into Eq. (16) yields:

$$Y_{lm}(\theta,\varphi) = \begin{cases} \sin m\lambda \sum_{i=0}^{N(l)} C_i \cos(l-2i)\theta \\ \cos m\lambda \sum_{i=0}^{N(l)} C_i \cos(l-2i)\theta \end{cases} \quad (19a)$$

for $m$ is even, and

$$Y_{lm}(\theta,\varphi) = \begin{cases} \sin m\lambda \sum_{i=0}^{N(l)} S_i \sin(l-2i)\theta \\ \cos m\lambda \sum_{i=0}^{N(l)} S_i \sin(l-2i)\theta \end{cases} \quad (19b)$$

for *m* is odd.

From Eq. (19) it is obvious that spherical harmonics are equivalent to a sum of terms of 2-D Fourier series. As a result, the temporal frequencies corresponding to spherical harmonics can be determined by the results in Subsection 3.1.

The temporal frequencies can be derived from Eq. (15) for the *l*-th degree spherical harmonics $\bar{Y}_{lm}(\theta,\lambda)$ when $q$ and $p$ are determined. Eq. (19) means that there are a sum of terms of the form $\sin m\lambda$ and $\cos m\lambda$ in the longitude direction, and $\sin(l-2i)\theta$ and $\cos(l-2i)\theta$ in the latitude direction. A comparison between Eq. (19) and Eq. (11) concludes that spatial frequencies in the longitude direction $p = 0, 1, 2, 3, …, l$, since the orders $m = 0, 1, 2, 3, …, l$ only; in the latitude direction $q = 0, 2, 4, …, l$ for *l* even, and $q = 1, 3, 5, …, l$ for *l* odd, since $i = 0, 1, 2, …, N(l)$. From $q = l - 2i$ it is also seen that $q$ and $l$ always have the same parity. Then the temporal frequencies for the *l*-th degree spherical harmonics $\bar{Y}_{lm}(\theta,\lambda)$ are obtained by using the Eq. (15). The temporal frequencies for $l = 0, 1, 2, 3, 4, 5$ are shown in the Table 1.

**Table 1** Temporal frequencies for spherical harmonics $\bar{Y}_{lm}(\theta,\lambda)$

($l = 0, 1, 2, 3, 4, 5$)

| l=0 | |
|---|---|
| p \ q | 0 |
| 0 | $f_{0,0}$ |

(a)

| l=1 | | |
|---|---|---|
| p \ q | | 1 |
| 0 | | $f_{1,0}$ |
| 1 | | $f_{1,1}$ |

(b)

| l=2 | | |
|---|---|---|
| p \ q | 0 | 2 |
| 0 | $f_{0,0}$ | $f_{2,0}$ |
| 1 | $f_{0,1}$ | $f_{2,1}$ |
| 2 | $f_{0,2}$ | $f_{2,2}$ |

(c)

| l=3 | | |
|---|---|---|
| p \ q | 1 | 3 |
| 0 | $f_{1,0}$ | $f_{3,0}$ |
| 1 | $f_{1,1}$ | $f_{3,1}$ |
| 2 | $f_{1,2}$ | $f_{3,2}$ |
| 3 | $f_{1,3}$ | $f_{3,3}$ |

(d)

| l=4 | | | |
|---|---|---|---|
| p \ q | 0 | 2 | 4 |
| 0 | $f_{0,0}$ | $f_{2,0}$ | $f_{4,0}$ |
| 1 | $f_{0,1}$ | $f_{2,1}$ | $f_{4,1}$ |
| 2 | $f_{0,2}$ | $f_{2,2}$ | $f_{4,2}$ |
| 3 | $f_{0,3}$ | $f_{2,3}$ | $f_{4,3}$ |
| 4 | $f_{0,4}$ | $f_{2,4}$ | $f_{4,4}$ |

(e)

| l=5 | | | |
|---|---|---|---|
| p \ q | 1 | 3 | 5 |
| 0 | $f_{1,0}$ | $f_{3,0}$ | $f_{5,0}$ |
| 1 | $f_{1,1}$ | $f_{3,1}$ | $f_{5,1}$ |
| 2 | $f_{1,2}$ | $f_{3,2}$ | $f_{5,2}$ |
| 3 | $f_{1,3}$ | $f_{3,3}$ | $f_{5,3}$ |
| 4 | $f_{1,4}$ | $f_{3,4}$ | $f_{5,4}$ |
| 5 | $f_{1,5}$ | $f_{3,5}$ | $f_{5,5}$ |

(f)

Table 1 shows that parts of the temporal frequencies for the *l*-th degree spherical harmonics are the same as for lower degree spherical harmonics, such as both $f_{1,0}$ and $f_{1,1}$ for $l = 1$ and $l = 3$. In order to estimate the errors introduced by the *l*-th degree spherical harmonics, the temporal frequencies contained in the lower degree spherical harmonics have to be deducted from the *l*-th degree. From Table 1, it is indicated that the more temporal frequencies contained in the *l*-th degree spherical harmonics than in the lower degrees ($\leq l - 1$) are the frequencies in the last column ($q = l$) and last two rows ($p = l$-1, $l$), which are filled up in gray, and concluded as:

$$f_j : \begin{cases} f_{l,0}, f_{l,1}, f_{l,2}, \cdots f_{l,l} & (A) \\ f_{0,l-1}, f_{2,l-1}, f_{4,l-1}, \cdots f_{l-2,l-1} & (B) \\ f_{0,l}, f_{2,l}, f_{4,l}, \cdots f_{l-2,l} & (C) \end{cases} \quad (20a)$$

for *m* is even, and

$$f_j : \begin{cases} f_{l,0}, f_{l,1}, f_{l,2}, \cdots f_{l,l} & (A) \\ f_{1,l-1}, f_{3,l-1}, f_{5,l-1}, \cdots f_{l-2,l-1} & (B) \\ f_{1,l}, f_{3,l}, f_{5,l}, \cdots f_{l-2,l} & (C) \end{cases} \quad (20b)$$

for *m* is odd.

where A-class, B-class and C-class frequencies stand for the last column (q = $l$), and last two rows ($p = l-1$, $l$), respectively. There are ($2l + 1$) frequencies for $l$ even and $2l$ frequencies for $l$ odd. In addition, from Eq. (20) it is seen that every degree spherical harmonics have low frequency signals, whose frequencies range from close to zero for $l$ even and from close to one cycle-per-revolution (cpr) for $l$ odd. As a result, it is noted that all degrees spherical harmonic coefficients recovery is affected by low frequency noise.

After this being said the effect of low frequency noise in all degrees is shown experimentally. For this purpose, it is assumed altitude $h = 250$ km and mission length $T = 2$ months (a full repeat period), which determine the velocity $\omega$ and the longitude lines $K_\lambda$, respectively. The temporal frequencies introduced by $l = 4$ and $l = 5$ is computed according to Eq. (20), as show in Table 2.

**Table 2** Temporal frequencies $f_j$ for spherical harmonics of the degrees $l = 4$ and $l = 5$

| Degree | A-class ($\times 10^{-4}$ Hz) | | | | | | B-class ($\times 10^{-4}$ Hz) | | C-class ($\times 10^{-4}$ Hz) | |
|---|---|---|---|---|---|---|---|---|---|---|
| $l = 4$ | 7.448 | 7.450 | 7.452 | 7.454 | 7.456 | | $5.787 \times 10^{-3}$ | 3.730 | $7.716 \times 10^{-3}$ | 3.732 |
| $l = 5$ | 9.310 | 9.312 | 9.314 | 9.316 | 9.318 | 9.320 | 1.870 | 5.594 | 1.872 | 5.596 |

It is obvious that A-class frequencies are close to $l$ cpr (1 cpr $\approx 1.86 \times 10^{-4}$ Hz); but B-class and C-class contain the frequencies close to zero frequency and 2 cpr for $l = 4$, the frequencies close to 1 cpr and 3 cpr for $l = 5$. As a result, it is certain that the recovery of high degree spherical harmonic coefficients is also affected by low frequency noise.

With the combination of Eq. (20) and Eq. (10), the direct relationship between the PSD of the satellite measurement errors and the coefficients of the Earth gravity potential is analytically concluded, and written as:

$$\sigma_l = \Omega \left\{ \sqrt{\sum_j \int_{f_j - \Delta f/2}^{f_j + \Delta f/2} \left[ S_n(f) \right]^2 df} \right\} \tag{21}$$

with

$$f_j : \begin{cases} f_{l,0}, f_{l,1}, f_{l,2}, \cdots f_{l,l} & \text{(A)} \\ f_{0,l-1}, f_{2,l-1}, f_{4,l-1}, \cdots f_{l-2,l-1} & \text{(B)} \\ f_{0,l}, f_{2,l}, f_{4,l}, \cdots f_{l-2,l} & \text{(C)} \end{cases}$$

for *l* even,

$$f_j : \begin{cases} f_{l,0}, f_{l,1}, f_{l,2}, \cdots f_{l,l} & (A) \\ f_{1,l-1}, f_{3,l-1}, f_{5,l-1}, \cdots f_{l-2,l-1} & (B) \\ f_{1,l}, f_{3,l}, f_{5,l}, \cdots f_{l-2,l} & (C) \end{cases}$$

for *l* odd, and

$$f_{q,p} = \frac{\omega}{2\pi}\left(q + \frac{2p}{K_\lambda}\right)$$

## 4 Simulation and discussion

As mentioned before, conventional methods address the effect of measurement errors and estimate the resolution of gravity field models from a numerical point of indirect view, and generally treat errors as white noise. In this section, this study analytically investigates the problems based on the above results.

### 4.2 White noise versus colored noise

The noise performed by gradiometers is significantly larger at both ends of the spectrum. The high-frequency tail of the noise spectrum, which is higher than measurement bandwidth, is generally disregarded by using a low-pass filter (ESA 1999; Mayrhofer et al. 2010). The frequencies close to the zero frequency are not measurable at all for the limit of measurement duration. In order to evaluate the effect in the whole range, three noise models are introduced in this study. It is assumed that the gradiometer noise is white over the entire measurement spectrum with a level of about 20 mE/$\sqrt{Hz}$ for the white noise model (Rummel et al 2011). The characteristic of colored noise model 1 is smoothed approximation of 1/*f* behavior below 27 cpr (≈ 0.005 Hz), and almost flat spectrum in the measurement bandwidth, whose PSD defined by means of the analytic function:

$$S(f) = \frac{S_0}{1 - e^{-\left(\frac{f}{f_0}\right)^\alpha}} \qquad (28)$$

where $S_0$ = 20 mE/$\sqrt{Hz}$, $f_0 \approx 0.005$Hz and $\alpha$ = 2. It should be pointed out that the parameter $\alpha$ is assigned a value of 1 by Ditmar et al. (2003), but assigned a value of 2 in this study, so that the actual characteristics can be more perfectly recovered. Colored noise model 2 is the same as colored noise model 1, except that beyond 2 cpr a flat spectrum with an amplitude in the order of the 2 cpr level was assumed (Schuh 2002). Fig. 2 displays all the spectral noise models as well as the GOCE gradiometer error PSD of gravity gradiometry tensor $V_{zz}$, which is based on two months of GOCE observations covering November and December 2009, i.e. a full repeat period.

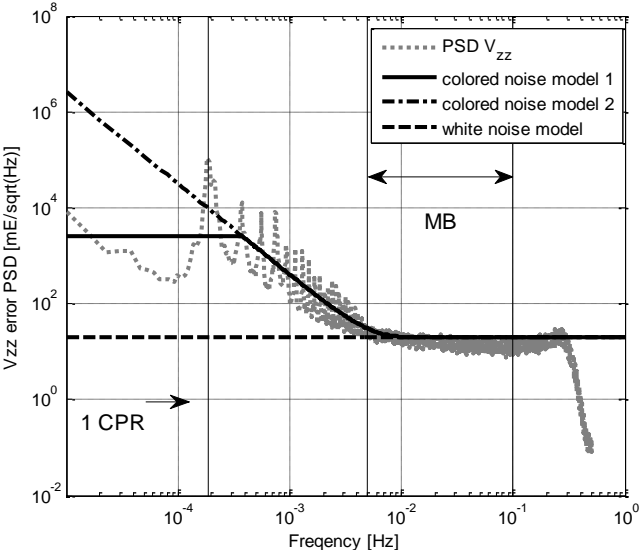

**Fig. 2** GOCE gradiometer error PSD of gravity gradiometry tensor $V_{zz}$ and theoretical noise models

Besides the assumptions mentioned above, the following parameters have been also used for the simulations: orbit height 250 km, mission duration 2 months and inclination 90°. Based on the results in Section 3, error degree power can be derived from Eq. (25), and error degree variance from Eq. (26), as shown in Figs. 3 and 4, respectively.

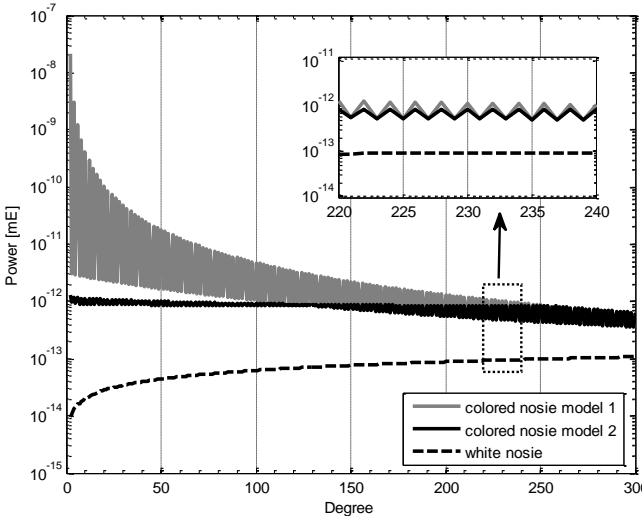

**Fig. 3** Error degree power with different model of gravity gradient measurement errors

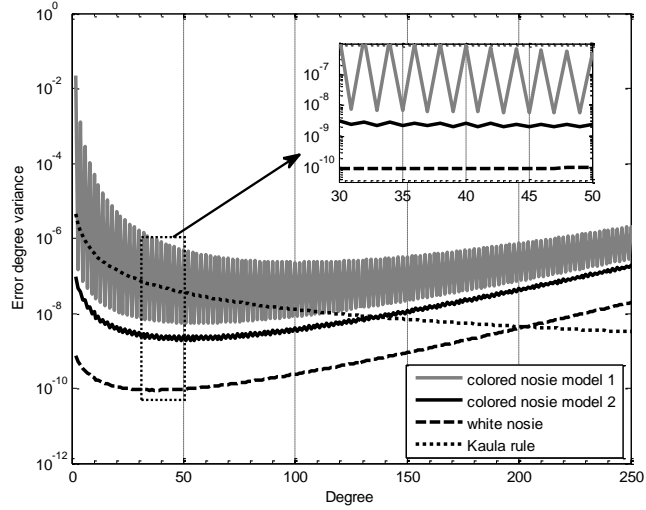

**Fig. 4** Error degree variance with different model of gravity gradient measurement errors

Based on Kaula's rule, it can be concluded that the maximum degree of the Earth's gravity field model is up to 203 from the white noise model, however, only 137 from colored noise model 2, and almost no information from colored noise model 1, as represented in Figs. 3 and 4. Moreover, the error degree variance of colored noise models is higher about 2~3 orders of magnitude than the white noise model in whole range. It shows that the gradiometry $1/f$ colored noise in the low-frequency range will seriously degrade the accuracy and resolution in all degrees. As mentioned in Subsection 3.2, the reason is that every degree spherical harmonics have low frequency signals.

Another significant phenomenon is that error degree variance is obviously a saw-tooth curve which fluctuates up and down depending on parity of $l$, crests for $l$ even and troughs for $l$ odd. The result is also appeared in the works by Rummel et al (1993) and Migliaccio et al.(2004), which analyze gravity field recovery with colored noise. This is also can be explained by the results in Subsection 3.2. When colored noise model applied, the error degree power is determined not only by the number of frequencies, but also the location of the frequencies in the spectrum, i.e. the values of the frequencies. Eq. (20), Tables 1 and 2 show that the spherical harmonics contain the frequencies close to the zero frequency for $l$ even, e.g. $f_{0,l}$ and $f_{0,l-1}$, as well to one cpr frequency for $l$ odd, e.g. $f_{1,l}$ and $f_{1,l-1}$. The PSD of $1/f$ colored noise is higher several orders of magnitude at the frequencies close to the zero frequency than other frequencies, so when *l* is even, the power introduced by *l*-th degree harmonics are larger than the (*l*-1)-th and (*l*+1)-th degrees. As the noise is white noise, the power introduced by *l*-th degree is only determined by the number of the frequencies, which is monotonically increasing function of *l*, therefore, the error degree variance under the condition of white noise is a smooth curve.

### 4.3 Influences of filters

Any data analysis from a gravity field mission has to apply a proper filter to obtain the available information with high efficiency (see Pail et al. 2007; Pail et al. 2011). For the purpose of exploiting the influence of filters, here three band-pass filters are employed on colored noise model 2: filter 1 with a bandwidth 0.8 mHz - 0.1Hz (~4 cpr - 537 cpr), filter 2 with a bandwidth 2mHz - 0.1Hz (~11 cpr - 537 cpr), and filter 3 with a bandwidth 5mHz - 0.1Hz (~27 cpr - 537 cpr), i.e. the interested measurement bandwidth of GOCE, as shown in Fig. 5.

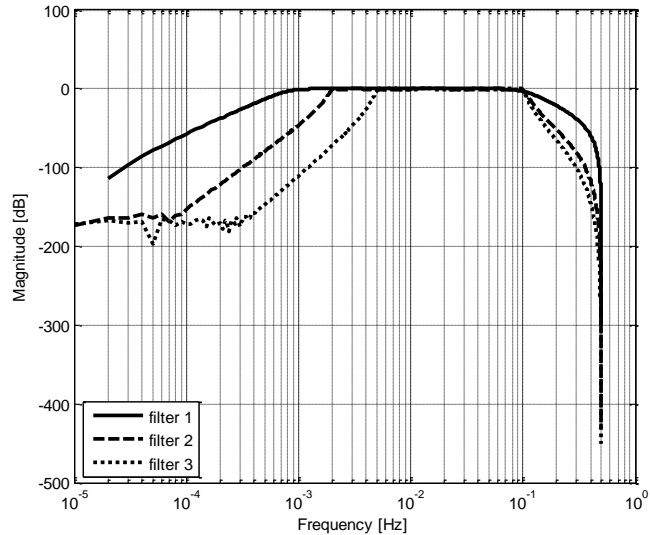

**Fig. 5** Magnitude-frequency characteristics of band-pass filters

The noise without measurement bandwidth is obviously attenuated by the filters. Then the error degree power and the error degree variance for the filtered data are derived, as shown in Figs. 6 and 7, respectively.

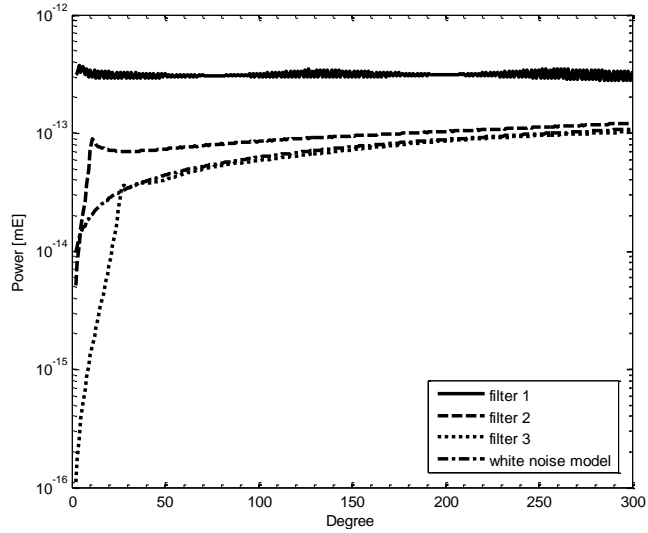

**Fig. 6** Error degree power with different filtered models of gravity gradient measurement errors

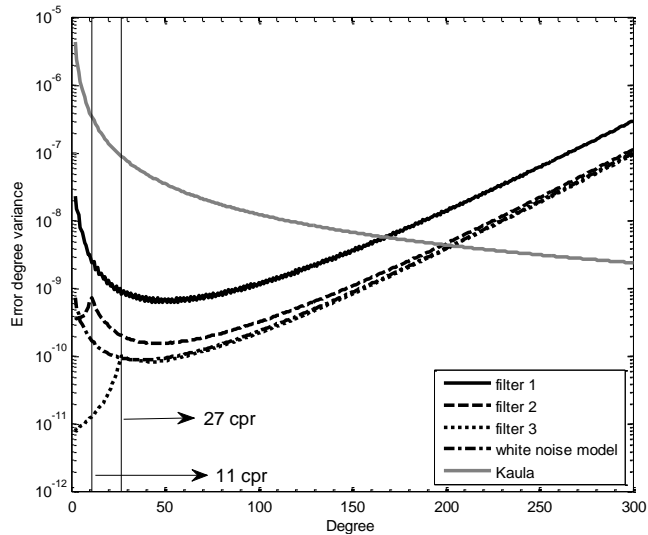

**Fig. 7** Error degree variance with different filtered models of gravity gradient measurement errors

Figures 6 and 7 show that the influence of $1/f$ colored noise is decreased caused by the reduction of the bandwidth of filters. The maximum recovery degrees of the gravity models applied the filters are 168, 199 and 204, respectively. Due to the lower cutoff frequencies of filters 2 and 3, which are 11 cpr and 27 cpr, error degree variance is obviously attenuated in the range of degrees $l \leq 11$ and $l \leq 27$, respectively. Fig. 7 shows that the error degree variance is reduced in the whole range for the reason that the colored noise in the low-frequency is filtered out. Certainly, the effect at lower degrees caused by the $1/f$ behavior is more obvious than higher degrees since the number of high frequencies increases with degrees.

It is also noted that the rest in the measurement bandwidth is almost white noise when the $1/f$ behavior below 5 mHz is filtered out by filter 3. Then, the error degree power is mainly determined by the number of the frequencies above 27 cpr, which is more than a half of the total number at least and increases with degrees. For these reasons, the error degree variance applied filter 3 and white noise model is almost the same in the range of degrees $l \geq 27$. However, it should be pointed out that the signals of gravity information at low frequencies are also filtered out when filters are employed. This will induce a problem that all coefficients of the gravity field model, which are only derived from high-pass filtered gradiometric data, are distorted, since that every degree spherical harmonics have low frequency signals. So the question, whether spherical harmonic coefficients are or not the best parameters to represent the potential, is always debatable (Sansò and Tscherning 2003). In order to solve this problem, a probably effective method for recovering the coefficients of the Earth's gravity field is to combine data from the gradiometric measurements and other kinds of observations, such as SST, satellite altimetry and terrestrial measurements, which are sensitive to the low-to-medium frequency constituents of the Earth's gravity field without a filter to reduce the low frequency signals. However, it still needs careful study.

## 5 Conclusions

Based on the 2-D Fourier theory and modulation theorem, the direct relationship between spherical harmonics and measurement errors is concluded. This study also exploits the effect of colored noise and filters on data analysis by taking advantage of the analytical expression. The method is verified to be an efficient and convenient procedure, especially for high accuracy and resolution gravity field models. Moreover, the results indicate that the low-frequency noise degrades the gravity field recovery in all degrees, so the colored noise must be processed carefully.

It should be noted that this study is based on the hypothesis that the satellite orbit is a polar circular orbit and observations acquired along the lines of longitude, i.e. ignoring the Earth rotation and related effects. Notwithstanding its limits, the essential relationship is clearly indicated. In addition, only the gravity gradient component $T_{zz}$ has been dealt with in this study. although the method can be also applied to other kinds of observables. Consequently, the other kinds of gradient components, high-low satellite-to-satellite tracking (hl-SST) and low-low satellite-to-satellite tracking (ll-SST) need to be further analyzed.

*Acknowledgements.* The authors are grateful to Prof. R. Pail (IAPG Munich) for his support of GOCE data, and Prof. Zhong Min (IGG Wuhan) for his discussions.# References

Cai L, Zhou Z, Zhu Z, Gao F, Hsu H (2011) Spectral analysis for recovering the Earth's gravity potential by satellite gravity gradient. Chinese J. Geophys. (in Chinese) (accepted)

Colombo OL (1981) Numerical methods for harmonic analysis on the sphere. Rep 310, Department of Geodetic Science, The Ohio State University, Columbus

Ditmar P, Kusche J, Klees R (2003) Computation of spherical harmonic coefficients from gravity gradiometry data to be acquired by the GOCE satellite: regularization issues. J Geod 77:465–477

ESA (1999) Gravity field and steady-state ocean circulation missions. Reports for mission selection. The four candidate Earth explorer core missions, SP-1233(1). European Space Agency, Noordwijk

Gruber C, Novák P, Sebera J (2011) FFT-based high-performance spherical harmonic transformation. Stud Geophys Geod, 55:489–500

Heiskanen W and Moritz H, 1967. Physical Geodesy. W. H. Freeman Company, San Francisco.

Ilk KH, Flury J, Rummel R, Schwintzer P, Bosch W, Haas C, Schröter J, Stammer D, Zahel W, Miller H, Dietrich R, Huybrechts P, Schmeling H, Wolf D, Riegger J, Bárdossy A, Güntner A (2004) Mass Transport and Mass Distribution in the Earth System, Contribution of the New Generation of Satellite Gravity and Altimetry Missions to Geosciences, Technische Universität München and GeoForschungsZentrum Potsdam

Klees R, Koop R, Visser P, van den IJssel J (2000) Efficient gravity field recovery from GOCE gravity gradient observations. J Geod, 74:561–571

Koop R (1993) Global gravity field modelling using satellite gravity gradiometry. Publ Geodesy, New Series, 38. Netherlands Geodetic Commission, Delft

Mayrhofer R; Pail R; Fecher T (2010) Quicklook gravity field solutions as part of the GOCE quality assessment. in: Lacoste-Francis, H. (eds.) Proceedings of the ESA Living Planet Symposium, ESA Publication SP-686, ESA/ESTEC, ISBN (Online) 978-92-9221-250-6, ISSN 1609-042X18